\documentclass[aps,prd,onecolumn,nofootinbib,superscriptaddress,showpacs,amsmath,amssymb]{revtex4-2}
\usepackage{graphicx,epsf}
\usepackage[]{latexsym}

\newcommand{\be}{\begin{eqnarray}}
\newcommand{\ee}{\end{eqnarray}}
\newcommand{\bea}{\begin{eqnarray}}
\newcommand{\eea}{\end{eqnarray}}

\def\comment#1{}

\usepackage{color}

\definecolor{darkred}{rgb}{.8,0,0}

\definecolor{darkblue}{rgb}{0,0,.7}

\definecolor{darkgreen}{rgb}{0,.7,0}

\begin{document}

\title{
Tsallis cosmology and its applications in dark matter physics with focus on IceCube high-energy neutrino data
}

\author{P. Jizba}
\email{p.jizba@fjfi.cvut.cz}
\affiliation{FNSPE, Czech Technical University in Prague, B\v{r}ehov\'{a} 7, 115 19 Praha 1, Czech Republic\\ITP, Freie Universitat Berlin, Arnimallee 14, D-14195 Berlin, Germany.}

\author{G. Lambiase}
\email{lambiase@sa.infn.it}
\affiliation{Dipartimento di Fisica "E.R. Caianiello", Universita' di Salerno, I-84084 Fisciano (Sa), Italy \&\\
INFN - Gruppo Collegato di Salerno, Italy}

%
%
%

\renewcommand{\theequation}{\thesection.\arabic{equation}}
\date{\today}
\begin{abstract}
\par\noindent
In this paper we employ a recent proposal of C.~Tsallis and formulate the first law of thermodynamics for gravitating systems in terms of the extensive but non-additive entropy. We pay a particular attention to an integrating factor for the heat one-form and show that in
contrast to conventional thermodynamics it factorizes into thermal and entropic part. Ensuing first law of thermodynamics implies Tsallis cosmology, which is then subsequently used to address the observed discrepancy between current bound on the Dark Matter relic abundance and present IceCube data on high-energy neutrinos. To resolve this contradiction 
we keep the conventional minimal Yukawa-type interaction between standard model and Dark Matter particles but replace  the usual Friedmann field equations with  Tsallis-cosmology-based modified Friedmann equations.
We show that when the Tsallis scaling exponent $\delta \sim 1.57$ (or equivalently, the holographic scaling exponent $\alpha \sim 3.13$) the aforementioned discrepancy disappears.
\end{abstract}
%

%
\maketitle

\section{Introduction}
\setcounter{equation}{0}
\label{intro}

In his seminal paper~\cite{Jacobson:1995ab}, Jacobson showed that there is a deep connection between gravity
and thermodynamics, with the possibility to derive the Einstein field equations from the first law of
thermodynamics (see also Refs.~\cite{Ver,Padmanabhan:2003gd,Padmanabhan:2009vy} for alternative approaches).
The important upshot of this approach is that one can infer the cosmological equations (the Friedmann equations) from the first law of thermodynamics
on the apparent horizon~\cite{Elin,Cai1,Pad,Cai2,Cai3,CaiKim,Fro,verlinde,Cai4,CaiLM,Shey1,Shey2}.
%
In recent years, there has been an upsurge of interest in extending this line of thoughts to cases where more general entropies than just the conventional Boltzmann--Gibbs entropy are employed~\cite{Tsallis:2012js,Komatsu:2013qia,Nunes:2014jra,Lymperis:2018iuz,Saridakis:2018unr,Sheykhi:2018dpn,Artymowski:2018pyg,Abreu:2017hiy,Jawad:2018frc,Zadeh:2018wub,daSilva:2018ehn}.
%
These models account for various
modifications of Bekenstein--Hawking's entropy area law. For instance, in the context of loop quantum gravity~\cite{Log,Rovelli,Zhang} or entropic cosmology~\cite{YFCai} the area law gets logarithmic corrections due to entanglement of quantum fields inside and outside the horizon~\cite{sau1,sau2,Sau,pavon1,lambiaseiorio}. Similarly, the generalized non-additive entropies~\cite{Tsa,Tsab,Jiz-a} often lead to more generic power-law instead of area law behavior.  A simple but prominent example of the latter is the so-called $\delta$-entropy
\begin{eqnarray}
S_{\delta}\ = \  \gamma_{\delta} A^{\delta}\, ,
\label{S}
\end{eqnarray}
where $A$ is the horizon area, $\delta$ is the scaling exponent and $\gamma_{\delta}$ is a $\delta$-dependent constant, which for $\delta=1$ reduces to Hawking's conventional form $\gamma=1/(4L_p^2)$. Entropy $S_{\delta}$ is a particular example of the entropy that was introduced by Tsallis in Ref.~\cite{Tsac} in order to define a correct thermodynamical entropy in 3 spatial dimensions for systems with the sub-extensive scaling, such as, e.g. black holes. This issue was then further elaborated, e.g., in Refs.~\cite{Tsa,Tsab}. Entropy $S_{\delta}$ is a special member of a two-parameter class of entropic functionals  known as $S_{q,\delta}$ (also introduced by Tsallis in~\cite{Tsa}), where  $S_{\delta} \equiv S_{1,\delta} $. It should be emphasized that $S_{\delta}$ has nothing to do with the popular {\em Tsallis entropy}~\cite{Tsallis:1987eu,Lyra:1998wz,Wilk:1999dr} that is widely used in statistical physics and theory of complex dynamical systems. Strictly speaking, Tsallis' entropy with the non-extensivity parameter $q$ is the $S_{q,1}$ member  in the previously mentioned two-parameter class of entropies.

The so-called {\em Tsallis cosmology} is a particular approach that accommodates the $S_{\delta}$ entropy directly into the first law of thermodynamics in order to arrive at the modified  cosmological Friedmann equations. The standard cosmological model is then recovered in the limit $\delta=1$. It should be noted that usage of $S_{\delta}$ in formulating the first law of thermodynamic seem a bit ad hoc in the current literature.  For this reason, we stick  in this paper to the original Tsallis proposal~\cite{Tsa} for $S_{\delta}$ and formulate the first law so that the entropy will be extensive but not additive. Due to a non additive nature of the entropy a particular attention must be paid to the integration factor of the heat one-form, which in this case is not a simple inverse of thermodynamic temperature, but instead it factorises into entropic and thermal part.

With the first law of thermodynamics at hand one can consider potential implications of the ensuing Tsallis cosmology. In particular, here we
show that the Tsallis cosmology is capable to alleviate the discrepancy between the current bound on the PeV Dark Matter (DM) relic abundance and recent IceCube data~\cite{2a,3a} about neutrino events with high energies ($\sim 1 $~PeV)~\cite{3a}.
Though astrophysical sources are favorite candidates for the origin of these high energy events~\cite{14a,15a,16a,Sahu:2014fua},
another viable possibility is that these neutrinos are the product of the decay of PeV mass Dark Matter~\cite{Aartsen:2016oji, 17a,21a,22a,23a,24a,25a,26a,27a,28a,29a,30a,31a,32a,33a,34a,35a,36a,37a} (see also Refs.~\cite{Griest:1989wd,Beacom:2006tt,40a,41a,52a,Griest:1989wd,merle,lambIce}). In this latter case,
the minimal extension of Standard Model (SM) describing such a decay is given by the renormalizable, dimension four (in mass units) Yukawa-type  interaction
 \begin{equation}\label{4dimop}
 {\cal L}_{4{\rm{-dim}}} \ = \ y_{\sigma \chi} \, \bar L_\sigma \cdot H \chi\,.
  \end{equation}
Here $\sigma = e, \mu, \tau$ indicates the  mass eigenstates of the three active neutrinos, $H$ the Higgs doublet, $L_{\sigma}$ the left-handed lepton doublet, $\chi$ the DM particle, and $y_{\sigma \chi}$ the (dimensionless) Yukawa coupling constants.
We calculate the freeze-in abundance of DM in Tsallis' cosmology and fix the parameter $\delta$ so as to
be consistent with the observed relic abundance and the IceCube requirements.
More precisely, the IceCube high energy events and the DM relic abundance are not compatible with the DM production if the latter is ascribed to the 4-dimensional operator (\ref{4dimop}).
This apparent tension can be resolved if one assumes that the Universe evolves according to Tsallis-cosmology implied Friedmann equations, provided one properly constrains the scaling exponent $\delta$.

The layout of the paper is as follows. In the next section we
discuss the role of $S_{\delta}$ entropy and place it in a proper thermodynamic framework. A particular attention is paid to an integrating factor for the heat one-form. It is  show that the latter cannot be simply identified with inverse thermodynamic temperature, but instead it factorises into entropic an thermal part. With the first law of thermodynamics obtained we
discuss in Section~\ref{FIRST} the ensuing modified Friedmann equations, which are obtain when one applies the first law,
at apparent horizon of a FRW (Friedmann--Robertson--Walker) Universe.
In Section \ref{sec3} we show that the DM relic abundance and the IceCube data can be explained in a consistent way by using the minimal 4-dimensional Yukawa-type interaction operator (\ref{4dimop}) and by assuming that the cosmological background is described via Tsallis cosmology.
Finally, Section~\ref{Conclusions} summarizes our results
and discusses possible extensions. For
the reader's convenience the paper is supplemented with one appendix which
clarifies more technical aspects associated to discrepancy between PeV neutrinos and IceCube data in the conventional FRW cosmology.

\section{Thermodynamic framework for the $S_{\delta}$ entropy}
\label{FIRSTA}
\setcounter{equation}{0}

Application of laboratory thermodynamics in (self-)gravitating systems is fraught with peril: from a formal point of view, it cannot even be defined, because there is no thermodynamic limit. The technical reason why one cannot scale the system to an infinitely large size stems from the fact that gravitation is a long-range force, which implies that the gravitational potential energy grows faster than a linear function of the mass of the system.

In many cosmological systems, such as black holes, this is typically rectified by employing entropies that have in 3 spatial dimensions a sub-extensive scaling (such as area law scaling of the Hawking--Bekenstein (HB) entropy). Recently Tsallis proposed~\cite{Tsa,Tsab} an alternative viewpoint, namely that such systems might still allow for a conventional thermodynamic description provided the entropy involved is extensive but not additive. In this section we will consider this proposal more seriously and compute the ensuing integration factor for the heat one-form.
We show that this not only allows to define the temperature but it will also provide a heuristic justification for the entropic form (\ref{S}).

We start by observing that the key property in thermodynamic framework is the Legendre transform which, for instance, for Gibbs free energy takes the form
\begin{eqnarray}
G(T,p,N, \ldots) \ = \ U(S,V,N, \ldots) \ + \ pV \ - \ T S \, ,
\label{B.1.c}
\end{eqnarray}
where $G$ and $U$ stand for Gibbs free energy and internal energy, respectively. Both $G$ and $U$ are expressed in terms of their {\em natural variables} and  dots stand for prospective additional state variables.

By following~\cite{Tsa,Tsab}, we now define the length-scale independent thermodynamic potentials $g = \lim_{L \rightarrow \infty} G/L^{\varepsilon}$ and  $u = \lim_{L \rightarrow \infty} U/L^{\varepsilon}$, where $L$ is the characteristic linear scale of the system and $\varepsilon$ is a scaling exponent (not necessarily identical with the spatial dimension $d$). Note that $g$ and $u$ must satisfy (for large $L$)
\begin{eqnarray}
&&G(T,p,N, \ldots) \ = \ L^{\varepsilon} g(T/L^{\theta},p/L^{\theta},N/L^d,\ldots ) \, ,\nonumber \\[2mm]
&&U(S,V,N, \ldots) \ = \ L^{\varepsilon} u(S/L^{d},1, N/L^d, \ldots)\, .
\label{B.2.c}
\end{eqnarray}
Here we do not assume that the scaling exponent $\theta$ has a typical laboratory value $\theta = 0$. So, because (for large $L$)   $G(T,p,N, \ldots) \propto L^{\varepsilon}$, $U(S,V,N, \ldots) \propto L^{\varepsilon}$,  $p \propto L^\theta$ and $T \propto L^\theta$, then (\ref{B.1.c}) inevitably implies that $S \propto L^d$ (this was implicitly used in (\ref{B.2.c})) and $\varepsilon = \theta + d$. In this way one can (for large $L$) rewrite (\ref{B.1.c}) in the form
\begin{eqnarray}
g(T/L^{\theta},p/L^{\theta},N/L^d,\ldots ) \ = \ u(S/L^{d},1, N/L^d, \ldots) \ + \ \frac{p}{L^{\theta}} \cdot 1 \ - \ \frac{T}{L^{\theta}} \frac{S}{L^d}\, .
\label{II.14.cc}
\end{eqnarray}
hence, the structure of Legendre transform is satisfied also for  length-scale independent thermodynamic potentials. A lesson that can be drawn from this analysis is that entropy should be  extensive quantity (provided that $T$ and $p$ scale in the same way) irrespective of the actual scaling of  thermodynamic potentials (which should be inevitably the same for all of them).
It is clear that one could repeat the same argument for other thermodynamic potentials.

Holographic principle posits that entropy of a black hole and more generally the entropy of the Universe is a Shannon entropy with a peculiar area-law scaling, namely
\begin{eqnarray}
S_{{\rm{HB}}} \ \propto \ - \sum_i p_i\log p_i \ = \ L^2\, ,
\label{II.15.cc}
\end{eqnarray}
where $L$ is a characteristic length-scale in the problem.
By the asymptotic equipartition property~\cite{Cover} is
\begin{eqnarray}
S_{{\rm{HB}}} \ \propto \ \log W\,,
\end{eqnarray}
where $W$ is a number of states (more precisely a volume of a typical set). So, $W$ should scale exponentially so that
\begin{eqnarray}
W \ = \ \phi(L) \eta^{L^2}\, , \;\;\; \mbox{with} \;\;\; \eta >1 \;\;\; \;\;\; \mbox{and} \;\;\; \lim_{L \rightarrow \infty} \phi(L)/L \ = \ 0\, .
\label{B.6.cf}
\end{eqnarray}

As argued before, the scaling of $S_{{\rm{BH}}}$  prevents to consider it as a full-fledged thermodynamic entropy. In this connection, it was argued in~\cite{Tsallis:2012js} that the entropy
\begin{eqnarray}
S_\delta \ \propto \ \sum_i p_i \left( \log \frac{1}{p_i}\right)^{\!\!\delta}\, , \;\;\;\; \delta \ > \ 0\, ,
\end{eqnarray}
which for equiprobable distribution behaves as~\footnote{Note that the positivity of $\delta$ implies that $S_\delta$ grows with an increasing number of available microstates. This might be viewed as a consistency condition on the validity of the ``generalized'' second law of thermodynamics. }
\begin{eqnarray}
S_\delta \ \propto \ \left(\log W \right)^{\delta} \, ,
\end{eqnarray}
might potentially represent a correct thermodynamic entropy in 3 spatial dimensions for systems with the sub-extensive scaling  (\ref{B.6.cf}).
Though many other non-additive and potentially useful entropies are available in the literature, cf. e.g.~\cite{E1,E2,E3,E4,PJ1,PJ2} our focus here will be on $S_{\delta}$. Among other, this will allow us to make contact with other results associated with Tsallis cosmology.

It is well known that the area-law formula for black hole entropy (\ref{II.15.cc}) holds only in Einstein theory,
i.e., when the ensuing action functional includes only a linear term of scalar curvature $R$. On the other hand, the area-law formula of black hole entropy no longer holds in generic higher-derivative gravity
theories~\cite{CaiKim}, for instance in $f(R)$ gravity the entropy of a static black hole acquires the form $S \propto L^2f'(R)$, cf. e.g.~\cite{Capo}. It is thus intriguing to consider entropy (\ref{II.15.cc}) with more general scaling law, namely
\begin{eqnarray}
S_{{\rm{Gen.HB}}} \ \propto \  L^{\alpha}\, ,
\end{eqnarray}
in other words, we allow for a deformation in the holographic scaling. So, $\delta$ from (\ref{S}) equals to $\alpha/2$. In the spirit of Tsallis suggestion we now assume that $S_{3/\alpha}$ is a thermodynamic entropy. There are two immediate impediments associated with this assumption. First, $S_{3/\alpha}$ is not additive (not even in the $L\rightarrow \infty$ limit) but it satisfies the pseudo-additivity  rule
\begin{eqnarray}
S_{3/\alpha}(A + B) \ = \ \left[S_{3/\alpha}^{\alpha/3}(A) \ + \  S_{3/\alpha}^{\alpha/3}(B) \right]^{3/\alpha}\, ,
\label{B.10.cf}
\end{eqnarray}
for any two independent subsystems $A$ and $B$. This is an inevitable consequence of working with systems with the sub-extensive scaling --- such as gravity. Second, it is not clear what is a thermodynamic conjugate to such an entropy.
Caratheodory theorem~\cite{Caratheodory,CaratheodoryII} ensures that heat one-form has an integration factor but since the entropy is not additive one cannot use the conventional Carnot cycle argument~\cite{Huang} in the proof of Clausius equality, to simply identify the integration factor with inverse temperature.

Let us dwell a bit more on this last point. Since the exact differential associated with the heat one-form
is entropy, we can write
\begin{eqnarray}
dS_{3/\alpha}({\bf{a}},\theta) \ = \ \mu({\bf{a}}, \theta) \delta Q({\bf{a}},\theta)\, ,
\end{eqnarray}
where ${\bf{a}}$ represent a collection of relevant state variables and $\theta$ is some {\em empirical} temperature whose existence is guaranteed by the zeroth law of thermodynamics.
We now divide the system in question into two subsystems $A$ and $B$, that are respectively described by state variables
$\{{\bf{a}}_1,\theta\}$ and $\{{\bf{a}}_2,\theta\}$, respectively. Then
\begin{eqnarray}
\delta Q_A({\bf{a}}_1,\theta) \ = \ \frac{1}{\mu_A({\bf{a}}_1,\theta)} \ \! dS_{A,3/\alpha}({\bf{a}}_1,\theta)\;\;\; \mbox{and} \;\;\; \delta Q_B({\bf{a}}_2,\theta) \ = \ \frac{1}{\mu_B({\bf{a}}_2,\theta)} \ \! dS_{B,3/\alpha}({\bf{a}}_2,\theta)\, .
\end{eqnarray}
So, for the whole system
\begin{eqnarray}
\delta Q_{A+B} \ = \ \delta Q_A \ + \ \delta Q_B \;\;\; \mbox{with} \;\;\; \delta Q_{A+B}({\bf{a}}_1,{\bf{a}}_2, \theta) \ = \ \frac{1}{\mu_{A+B}({\bf{a}}_1,{\bf{a}}_2,\theta)} \ \! dS_{(A+B),3/\alpha}({\bf{a}}_1,{\bf{a}}_2,\theta)\, ,
\end{eqnarray}
we can write
\begin{eqnarray}
dS_{(A+B),3/\alpha}({\bf{a}}_1,{\bf{a}}_2,\theta) \ = \ \frac{\mu_{A+B}({\bf{a}}_1,{\bf{a}}_2,\theta)}{\mu_A({\bf{a}}_1,\theta)} \ \! dS_{A,3/\alpha}({\bf{a}}_1,\theta) \ + \
 \frac{\mu_{A+B}({\bf{a}}_1,{\bf{a}}_2,\theta)}{\mu_B({\bf{a}}_2,\theta)} \ \! dS_{B,3/\alpha}({\bf{a}}_2,\theta)\, .
 \label{B.14.fg}
\end{eqnarray}
Let us now assume that there is only one state variable (apart from $\theta$), so that ${\bf{a}} = a$. If that there would be more state variables, our following argument would go through as well but we would need to employ more than two subsystem. Under this assumption we can invert $S_{A,3/\alpha}(a_1,\theta)$ and $S_{B,3/\alpha}(a_b,\theta)$  and write
\begin{eqnarray}
a_1 \ = \ a_1(S_{A,3/\alpha}, \theta) \;\;\; \mbox{and} \;\;\; a_2 \ = \ a_2(S_{B,3/\alpha}, \theta) \, .
\end{eqnarray}
With this  (\ref{B.14.fg}) can be rewritten as
\begin{eqnarray}
dS_{(A+B),3/\alpha}(S_{A,3/\alpha},S_{B,3/\alpha},\theta) \ &=& \ \frac{\mu_{A+B}(S_{A,3/\alpha},S_{B,3/\alpha},\theta)}{\mu_A(S_{A,3/\alpha},\theta)} \ \! dS_{A,3/\alpha} \nonumber \\[2mm]
&+& \
 \frac{\mu_{A+B}(S_{A,3/\alpha},S_{B,3/\alpha},\theta)}{\mu_B(S_{B,3/\alpha},\theta)} \ \! dS_{B,3/\alpha} \ + \ 0 \ \! d\theta\, .
 \label{B.16.cb}
\end{eqnarray}
Since $dS_{3/\alpha}$ (for all considered systems) must be a total differential (so that $S_{3/\alpha}$ is a proper state function), integrability conditions give
\begin{eqnarray}
&&\frac{\partial \log(\mu_A(S_{A,3/\alpha},\theta))}{\partial \theta} \ = \ \frac{\partial \log(\mu_B(S_{B,3/\alpha},\theta))}{\partial \theta} \ = \ \frac{\partial \log(\mu_{A+B}(S_{A,3/\alpha},S_{B,3/\alpha},\theta))}{\partial \theta}\, , \label{B.17a.cv} \\[2mm]
&& \frac{1}{\mu_A(S_{A,3/\alpha},\theta)} \frac{\partial \mu_{A+B}(S_{A,3/\alpha},S_{B,3/\alpha},\theta)}{\partial S_{B,3/\alpha}} \ = \ \frac{1}{\mu_B(S_{B,3/\alpha},\theta)} \frac{\partial \mu_{A+B}(S_{A,3/\alpha},S_{B,3/\alpha},\theta)}{\partial S_{A,3/\alpha}}\, . \label{B.17b.cv}
\end{eqnarray}
Note that in (\ref{B.17a.cv}) the derivatives cannot depend on entropy but only on $\theta$. By denoting the RHS of (\ref{B.17a.cv}) as $-\omega(\theta)$
we might resolve (\ref{B.17a.cv}) in the form
\begin{eqnarray}
&&\mu_A(S_{A,3/\alpha},\theta) \ = \ \psi_A(S_{A,3/\alpha}) \ \!\exp\left(-\int \omega(\theta) d\theta \right)\, ,\nonumber \\[2mm]
&&\mu_B(S_{B,3/\alpha},\theta) \ = \ \psi_B(S_{B,3/\alpha}) \ \!\exp\left(-\int \omega(\theta) d\theta \right)\, ,\nonumber \\[2mm]
&&\mu_{A+B}(S_{A,3/\alpha},S_{B,3/\alpha},\theta)) \ = \ \psi_{A+B}(S_{A,3/\alpha},S_{B,3/\alpha}) \ \!\exp\left(-\int \omega(\theta) d\theta \right)\,,
\label{B.19.hh}
\end{eqnarray}
where $\psi$ are some arbitrary functions of the entropy. Note that the temperature part of $\mu$ is $\alpha$ independent.

Let us now observe from (\ref{B.10.cf}) that
\begin{eqnarray}
dS_{(A+B),3/\alpha} \ = \ \frac{S_{A,3/\alpha}^{\alpha/3-1}}{S_{(A+B),3/\alpha}^{\alpha/3-1}}   \ \! dS_{A,3/\alpha} \ + \ \frac{S_{B,3/\alpha}^{\alpha/3-1}}{S_{(A+B),3/\alpha}^{\alpha/3-1}}   \ \! dS_{B,3/\alpha}\, .
\label{II.20.cf}
\end{eqnarray}
By comparing this with (\ref{B.16.cb}) and (\ref{B.19.hh}) with (\ref{II.20.cf}) we can make identification $\psi(S_{\bullet, 3/\alpha}) = \varsigma S^{1-\alpha/3}_{\bullet,3/\alpha}$, where $\varsigma$ is a constant and $\bullet$ stands for $A$, $B$ and $A+B$, respectively. With this identification one can easily check that also the integrability condition
(\ref{B.17b.cv}) is satisfied. In conventional  thermodynamics $\psi$ would be only a constant and so the integration factor could be identified with a genuine absolute temperature. In the context of non-additive entropy  $S_{3/\alpha}$ we see that this is not so. Fortunately $\mu$ has a simple factorized form.

Let us now call the temperature part in $\mu$ in (\ref{B.19.hh}) as $1/T$.
So, the corresponding thermal contribution that enters the first thermodynamic law will have the form
\begin{eqnarray}
\frac{1}{\mu} \ \!d S_{3/\alpha} \ = \ T \frac{S^{\alpha/3-1}_{3/\alpha}}{\varsigma} \ \!d S_{3/\alpha} \ = \ \frac{T}{\varsigma\alpha/3} \ \! d  S_{3/\alpha}^{\alpha/3}\, .
\end{eqnarray}
In the following we will denote $3S_{3/\alpha}^{\alpha/3}/\varsigma\alpha$  as ${\mathcal{S}}_{(\alpha)}$.  Note that ${\mathcal{S}}_{(\alpha)} \propto L^{\alpha}$ and so it behaves in the same way as $S_{\alpha/2}$ in Eq.~(\ref{S}). By analogy with~(\ref{S}) we set the proportionality factor to be $(4\pi)^{\alpha/2}\gamma_{\alpha/2}$, where $\gamma_{\alpha/2}$ still needs to be determined. Finally, we can write the first law of thermodynamics in the form
\begin{eqnarray}
dU\ = \ T d{\mathcal{S}}_{(\alpha)} \ - \ pdV\,,
\label{II.22.cv}
\end{eqnarray}
In passing we note that in the cosmology framework (similarly as, e.g. in fluid dynamics) the role of pressure is taken over by the {\em work density} $W$.

\section{Tsallis cosmology from the First law of thermodynamics\label{FIRST}}

We assume a homogeneous and isotropic Universe (i.e., FRW Universe), thus the line element is the  given by
$ds^2={h}_{\alpha \beta}dx^{\alpha} dx^{\beta}+\tilde{r}^2(d\theta^2+\sin^2\theta d\phi^2)$,
where $\tilde{r}=a(t)r$, $x^{\alpha}=(t, r)$, $\alpha=0, 1$, $h_{\alpha \beta}=(-1, a^2/(1-kr^2))$ is a two-dimensional metric, and  $k$ is a constant (spatial) curvature parameter with values $-1,0,+1$ representing open, flat and closed geometry, respectively.
The physical boundary of the Universe is assumed to be given by the apparent horizon with radius~\cite{Sheykhi}
\begin{eqnarray}
 \tilde{r}_A \ = \ \frac{1}{\sqrt{H^2+k/a^2}}\, ,
 \label{III.23.cd}
\end{eqnarray}
($H=\dot{a}/a$ is the Hubble parameter) with the associated surface gravity and
 horizon temperature~\cite{Cai2,Sheykhi}
\begin{eqnarray}
{\kappa} \ = \ \left. \frac{1}{2 \sqrt{-h }} \ \! \partial_{\alpha}\left( \sqrt{-h} h^{\alpha \beta} \partial_{\beta} \tilde{r}\right)\right|_{\tilde{r} = {\tilde r}_A}   \ = \  - \frac{1}{ \tilde
r_A}\left(1-\frac{\dot {\tilde r}_A}{2H\tilde r_A}\right)  \;\;\;\;\;\; \mbox{and } \;\;\;\;\; T \ = \ \frac{|\kappa|}{2\pi}\, .
\end{eqnarray}
Here $\dot {\tilde r}_A \equiv d{\tilde r}_A/dt$, $\kappa$ is the surface gravity and $H $ is the Hubble parameter. Usually one assumes that the apparent horizon radius is (almost) fixed~\cite{CaiKim,Sheykhi,cao},  which allows to set $\dot {\tilde r}_A\ll 2H\tilde r_A$. This implies that the volume (almost) does not change and one may thus simply take  $T= 1/(2\pi \tilde r_A )$.
In order to satisfy the field equations, the symmetries of the Einstein tensor, imply that the energy-momentum tensor  in the FRW Universe
must  have the perfect fluid form: $T_{\mu\nu}=(\rho+p)u_{\mu}u_{\nu}+pg_{\mu\nu}$, where $\rho$ and $p$ are the energy density and pressure, respectively, which are bound together via the continuity equation
\begin{eqnarray}
 \dot{\rho}\  + \ 3H(\rho \ + \ p) \ = \ 0\, .
 \label{III.25.cc}
\end{eqnarray}
The {\it work density} (which is due to the change in the apparent horizon radius), for a FRW Universe assumes the form $W=- \frac{1}{2}{\mbox{Tr}} (T^{\mu\nu}) =\frac{1}{2}(\rho-p)$ where ``$\mbox{Tr}$'' denotes the two-dimensional normal trace, i.e. $ {\mbox{Tr}} (T^{\mu\nu}) =  T^{\alpha \beta} h_{\alpha\beta}$.
The first law of thermodynamics~(\ref{II.22.cv}) then acquires the form
\begin{eqnarray}
dU\ = \ T d{\mathcal{S}}_{(\alpha)} \ - \ WdV\, .
\label{III.26.cf}
\end{eqnarray}
%
We now use the fact that the increase in internal energy $dU$ due to the change of the apparent horizon volume (i.e. a $3$-sphere of radius $\tilde{r}_{A}$) corresponds to the decrease in the
total energy content $E$ of the Universe inside of the volume, 
so that $dU =-dE$, and rewrite (\ref{III.26.cf}) in the form
\begin{eqnarray}
dE \ = \ - Td{\mathcal{S}}_{(\alpha)} \ + \ WdV\, ,
\label{III.27.cd}
\end{eqnarray}
which formally coincides with the so-called {\em the unified
first law}~\cite{Hayward}.  By employing the relation $E=\rho V$
(with $V=\frac{4\pi}{3}\tilde{r}_{A}^{3}$ being the apparent horizon volume)
%
and by setting $L$ in the definition of ${\mathcal{S}}_{(\alpha)}$ to be $\tilde{r}_{A}$, we obtain from (\ref{III.27.cd}) that
\begin{eqnarray}
V\dot{\rho} dt \ + \ \rho dV \ = \ - T d\left[ (4\pi)^{\alpha/2}\gamma_{\alpha/2} \ \! \tilde{r}_{A}^\alpha\right] \ + \ \frac{1}{2}(\rho - p)dV\, .
\end{eqnarray}
If we now employ (\ref{III.25.cc}), the fact that $\dot{\tilde{r}}_A/\tilde{r}_A \ll 2 H$ and identity $dV = 4\pi \tilde{r}_{A}^2 d \tilde{r}_{A}$, we arrive at
\begin{eqnarray}
\frac{\alpha}{2\pi \tilde{r}_{A}^3} \ \! \gamma_{\alpha/2} \ \! (4 \pi \tilde{r}_{A}^2)^{\alpha/2-1} \ \! d\tilde{r}_{A} \ = \  H (\rho + p) \ \! dt \ = \ - \frac{1}{3} \ \! d \rho\, .
\label{III.28.bb}
\end{eqnarray}
The last equality is due to (\ref{III.25.cc}). Eq.~(\ref{III.28.bb})
is nothing but a differential version of Friedmann equation. Ensuing differential equation can be solved yielding
\begin{eqnarray}
\rho(\tilde{r}_{A}) \ = \ \left( \tilde{r}_{A}^2 \right)^{\alpha/2 -2} \left[\frac{2  (4 \pi)^{\alpha/2 - 2} \ \! \alpha \ \! \gamma_{\alpha/2}}{4- \alpha}  \right] \ + \ c\,.
\end{eqnarray}
By requiring that the matter density is positive and that for $\tilde{r}_{A} \rightarrow \infty$ the density $\rho \rightarrow 0$, we see that $\alpha <4$,  the integration constant $c$ should be set to zero, and in addition $\gamma_{\alpha/2}>0$.
If we now employ (\ref{III.23.cd}) we obtain the {\em first modified Friedmann equation}
\begin{equation} \label{Fried4}
\frac{8\pi M_{\rm{Pl}}^{-\alpha}} {3} \rho \ = \ \left(H^2+\frac{k}{a^2}\right)^{2-\alpha/2}  \, ,
\end{equation}
where $\gamma_{\alpha/2}$ appearing in (\ref{S}) assumes the form
\begin{eqnarray}
\gamma_{\alpha/2}\ = \ \frac{3(4-\alpha)(4\pi)^{1-\alpha/2} M_{\rm{Pl}}^{\alpha}}{4\alpha }\, ,
\label{III.31.bb}
\end{eqnarray}
%
where $M_{\rm{Pl}}$ is the Planck mass.

The {\em second modified Friedmann equation} governing the evolution of the Universe in Tsallis cosmology
can be obtained by taking time derivative of (\ref{Fried4}) and using (\ref{III.25.cc}). With this we get, cf. also
~\cite{Sheykhi}
\begin{eqnarray}\label{2Fri5}
\frac{\ddot{a}}{a} \left(H^2 + \frac{k}{a^2} \right)^{1-\alpha/2} \ = \ \frac{8\pi M_{{\rm{Pl}}}^{-\alpha}}{3(4-\alpha)} \left[(1-\alpha) \rho - 3p\right]\,.
\end{eqnarray}
Cosmological observations such as Type Ia SNe~\cite{Riess}, CMB~\cite{Spergel} and the large scale structure~\cite{Tegmark,Eisenstein}, indicate that
the Universe is currently in an accelerated phase, hence $\ddot{a}>0$. By using the equation of state $p=\omega \rho$ we get from (\ref{2Fri5}) that
\begin{eqnarray} \label{w1}
(1-\alpha)\rho \ - \ 3\omega \rho \ > \ 0  \;\;\;  \Rightarrow \;\;\; \omega < (1-\alpha)/3  \,.
\end{eqnarray}
From this we see that for $\alpha \geq 1$ it always follows that $\omega< 0$, while for $\alpha <1$ one can have also  $\omega\geq 0$, thus
in Tsallis cosmology the accelerated phase of the late time Universe is possible even with the ordinary matter.
In particular, for $\omega=0$ (ordinary dust matter) the accelerated expansion can be obtained with the  scaling exponent  $\alpha < 1$.
Here, however, we will not explore this interesting issue any further.

\section{PeV neutrinos and DM relic abundance in Tsallis cosmology}
\label{sec3}
\setcounter{equation}{0}

The conventional  Yukawa-type  interaction  (\ref{4dimop})
fails to explain the IceCube data on high-energy neutrinos events ($\sim 1 $~PeV)
and PeV DM relic abundance  if the cosmological background evolves according to Einstein field equations (see Appendix, cf. also e.g.,~\cite{merle,lambIce}). We will now demonstrate that such a discrepancy can be avoided provided that the cosmological background evolves according to Tsallis cosmology.
To this end, we first cast the modified Friedmann  equation (\ref{Fried4}) in the form (we set $k=0$)
 \begin{equation}\label{H=AHGRIce}
   H(T)\ = \  Q(T)\,H_{{\rm{St.Cosm.}}}(T)\,,
 \end{equation}
where $H_{{\rm{St.Cosm.}}}=\displaystyle{\sqrt{\frac{8\pi}{3M^2_{{\rm{Pl}}}}\, \rho(T)}}$ is the Hubble parameter in the standard cosmology, and $Q(T)$ represents  the amplification factor, which in the present case reads
\begin{eqnarray}
  Q(T) \ &= & \ \left[\sqrt{\frac{8\pi}{3}}\, \frac{\rho^{1/2}}{M_{{\rm{Pl}}}^2}\right]^{\frac{\alpha-2}{4-\alpha}} \label{ATsallis}
  \ = \  \eta \left(\frac{T}{T_*}\right)^{\!\nu}\,,  \label{ATsallis2}  \\[2mm]
   \eta  \ & = & \ \left[\frac{2\pi}{3} \sqrt{\frac{\pi \ \! g_*(T)}{5}}\right]^{\frac{\alpha-2}{4-\alpha}}\,, \label{etaTsallis} \\[2mm]
  \nu & = & \frac{2(\alpha-2)}{4-\alpha} \,, \quad T_* \ = \  M_{{\rm{Pl}}}\,. \label{nuTsallis}
\end{eqnarray}
Here we have used the relation
$\rho=\frac{\pi^2 g_*(T)}{30}\,T^4$, with $g_*(T)\sim 106$ the effective number of degrees of freedom.
The equation of state for the radiation, $p=\rho/3$, implies that the continuity equation reads $\dot{\rho}(t)+4H\rho(t)=0$ from which one simply gets that $\rho(t)={\rho_0}/{a^4(t)}$, where $\rho_0$ is a constant. Inserting this expression for $\rho$ into (\ref{Fried4}), one obtains $a(t)= a_0 \left(\frac{t}{4-\alpha}\right)^{1-\alpha/4}$. The latter implies that there is a simple relation between the {\em cosmic time} and the {\em temperature}, namely ${t} \propto  T^{\frac{4}{\alpha-4}}$, which in turn gives that  the product $T a(t) = {\rm{constant}}$.

\begin{figure}[btp]
 \centering
  \includegraphics[width=10.0cm]{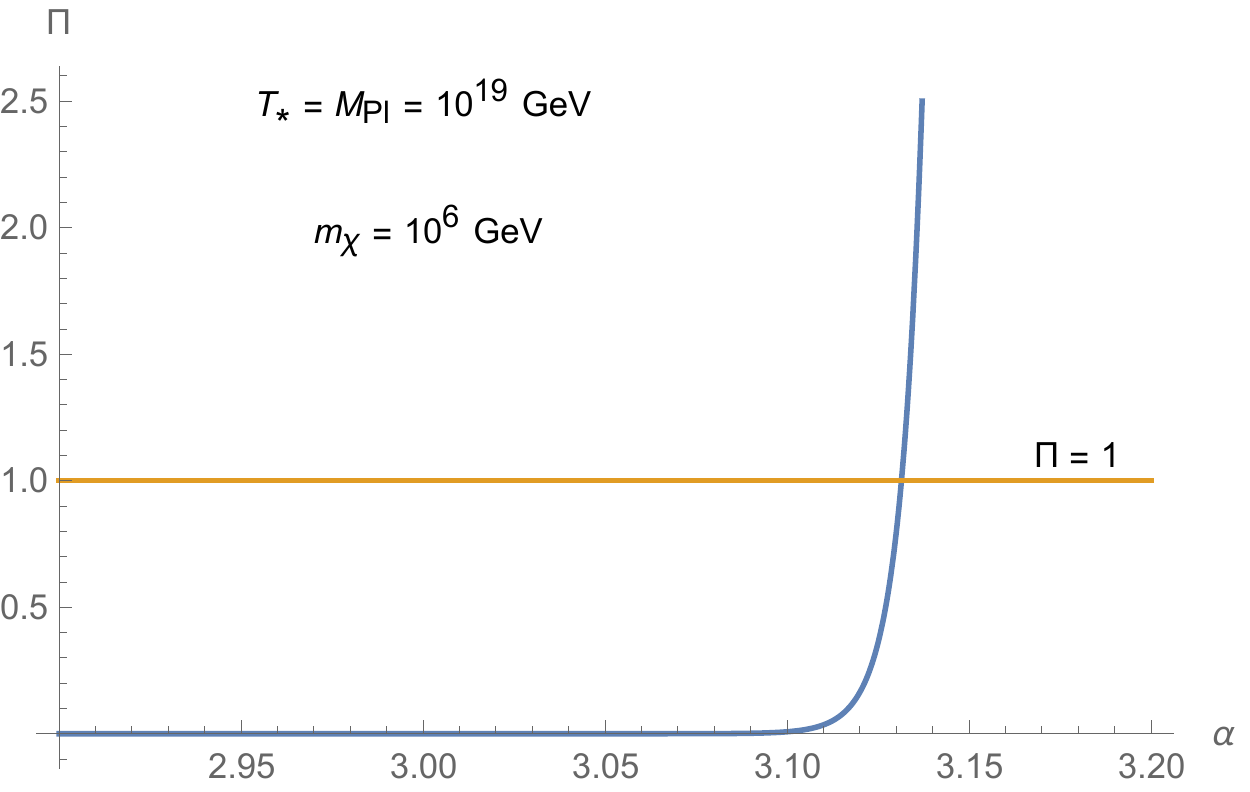}\\
  \caption{$\Pi$ vs $\alpha$}\label{Pi}
\end{figure}

The modified expansion rate (\ref{H=AHGRIce}) allows to write the inverse decay processes (see Eq.~(\ref{dYinvdec})) in the form
\begin{equation}\label{dYMC}
\frac{dY_\chi}{dT}\Big|_{{\rm{inv.dec.}}} \ = \ -\frac{m_\chi^2 \Gamma_\chi}{\pi^2 s H}\, K_1\left(\frac{m_\chi}{T}\right)\,,
\end{equation}
where
 \begin{equation}\label{Heta}
  s H \ = \ \frac{3.32\pi^2}{45}g_*^{3/2}\eta\left(\frac{m_\chi}{T_*}\right)^\nu \frac{1}{z^{5-\nu}}\,,
  \quad \quad z \ \equiv \ \frac{T}{T_*}\,.
 \end{equation}
The DM relic abundance follows by using\footnote{In this calculations we use $\displaystyle{\int_0^\infty dx x^{3+\nu}K_1(x)=2^{2+\nu}\Gamma\left(\frac{5+\nu}{2}\right)\Gamma\left(\frac{3+\nu}{2}\right)}$, where $\Gamma(z)$ are the Gamma functions. Notice also that
$\Gamma\left(\frac{5}{2}\right)\Gamma\left(\frac{3}{2}\right)=\frac{3\pi}{2}$.}
(\ref{dYMC}) and (\ref{Heta}) (see Eq.~(\ref{DM1}))
  \begin{eqnarray}\label{DM1Text}
   \Omega_{{\rm{DM}}}h^2 \ &=& \ \frac{2 m_\chi^2 s_0 h^2}{\rho_{{\rm{cr}}}}
   \int_0^\infty\frac{dx}{x^2}\left(-\frac{dY_\chi}{dT}\Big|_{T=\frac{m_\chi}{x}}\right)\,, \\[2mm]
   & = & \frac{45h^2}{1.66\pi^2 g^{3/2}}\frac{s_0 M_{Pl}}{\rho_{{\rm{cr}}}}\frac{\Gamma_\chi}{m_\chi}\frac{2^{2+\nu}}{\eta}\left(\frac{T_*}{m_\chi}\right)^\nu
   \Gamma\left(\frac{5+\nu}{2}\right)\Gamma\left(\frac{3+\nu}{2}\right) \label{DMeta} \nonumber \\[2mm]
  &\simeq & \ 0.1188 \left(\frac{106,7}{g_*}\right)^{3/2}\frac{\sum_\sigma |y_{\sigma\chi}|^2}{10^{-58}}\, \Pi\,,\nonumber
 \end{eqnarray}
where $\rho_{{\rm{cr}}}$ is the critical density, $s_0$ is the present value of the entropy density and $h$ is ``little $h$'', i.e. the dimensionless Hubble constant, and
%
 %
 %
 \begin{eqnarray}
  \Pi \ = \ 10^{-33}\, \frac{2^{\nu}}{7.5 \ \!\eta}\left(\frac{T_*}{m_\chi}\right)^\nu
   \frac{\Gamma\left(\frac{5+\nu}{2}\right)\Gamma\left(\frac{3+\nu}{2}\right)}{\Gamma\left(\frac{5}{2}\right)\Gamma\left(\frac{3}{2}\right)}\,. \label{Phi}
\end{eqnarray}
The DM relic abundance ($\Omega_{{\rm{DM}}}h^2 \sim 0.1188$) and the IceCube data ($\sum_\sigma |y_{\sigma\chi}|^2 \sim 10^{-58}$) can be consistently explained provided
 \begin{equation}\label{Phivalue}
   \Pi \ \simeq \ 1\,.
 \end{equation}
Using for the DM mass $m_\chi \sim \,\,\,$PeV $\sim 10^6$GeV and for $T_* = M_{{\rm{Pl}}} = 10^{19}$GeV, it then follows that
%
$\alpha  \simeq  3.131$.
%
A numerical solution is shown in Fig.~\ref{Pi}. Remarkably, such a value is $< 4$, as required by a positive matter-field density, and $> 0$ as needed for the validity of the generalized second law of thermodynamics.

\section{Conclusions}
\label{Conclusions}

To reconcile both the current bound on DM relic abundance and IceCube high-energy events of neutrinos
one must either change the minimal Yukawa-type interaction between standard model particles and DM particle or change the standard cosmological model or possibly modify both. Our modus operandi in this paper was to keep the
Yukawa-type interaction $\overline{L}_{\sigma}H\chi$ unchanged but instead replace the FRW cosmology
with the Tsallis cosmology and the Friedmann field equations with ensuing modified Friedmann equations. This proved to be a useful strategy.

In order to provide a sound thermodynamic basis for $S_{\delta}$ entropy and ensuing first thermodynamic law a particular attention has been paid to an integrating factor for the heat one-form. We have shown that the latter cannot be simply identified with inverse thermodynamic temperature, but instead it factorises into entropic an thermal part.

With the first law of thermodynamics at hand
we have shown that the Tsallis cosmology is capable to resolve the aforementioned discrepancy with only one additional adjustable parameter --- Tsallis scaling exponent $\alpha$. The main idea of our proof relies on the observation that in non-additive cosmology of Tsallis the expansion rate of the Universe can be cast in the form $H(T)=Q(T)H_{{\rm{St.Cosm.}}}(T)$, where $Q(T)$ encodes the parameters characterizing the model of gravity. Consequently, also the thermal history of Universe and DM in it are altered.
With the modified expansion rate we have solved the Boltzmann equation and obtained the abundance of DM particles.
Results attained are consistent with observed DM relic abundance provided $\alpha \sim 3.131$ (or equivalently $\delta \sim 1.565$), which is well in the range of validity of $\alpha \in (0,4)$.

\section{Acknowledgement}

GL thanks MUR and INFN for support. PJ was in part supported by the FNSPE CTU grant RVO14000.

\appendix


\section{PeV neutrinos and IceCube data in GR}
\label{AppPeV}

In this Appendix, we recall the main features related to DM relic abundance and IceCube data~\cite{merle,lambIce} by using conventional FRW cosmology.
The simplest 4-dimensional operator able to explain the IceCube high energy signal, is given by the Yukawa-type interaction (\ref{4dimop}).
%
 %

We consider the {\it freeze-in} production, that is the DM particles are never in thermal equilibrium owing to the fact that their interaction is very weak, and are produced from the hot thermal bath \cite{hall,merle} . 
%
As usual, the evolution of the DM particles is described by the Boltzmann equation. Defining $Y_\chi=n_\chi/s$ the DM abundance (here $n_\chi$ is the number density of the DM particles and $s=\frac{2\pi^2}{45}g_*(T)T^3$ the entropy density, with $g_*\simeq 106.75$ the degrees of freedom), the Boltzmann equation  gives \cite{merle}
 \begin{equation}\label{Boltz}
   \frac{dY_\chi}{dT} \ = \ -\frac{1}{H_{{\rm{St.Cosm.}}} T s}\left[\frac{g_\chi}{(2\pi)^3}\int C\frac{d^3p_\chi}{E_\chi}\right]\,,
 \end{equation}
where $H_{{\rm{St.Cosm.}}}$ is the expansion rate of the Universe described by conventional Friedmann equation (i.e., the Hubble parameter in the standard cosmology), $g_\chi$ is the number of degrees of freedom for DM (in our case it corresponds to two helicity projections, i.e. $g_\chi = 2$) and $C$ is the general collision term. In the case in which $dg_*/dT=0$, the DM relic abundance acquires the form \cite{merle}
 \begin{equation}\label{DM1}
   \Omega_{{\rm{DM}}}h^2 \ =  \ \frac{2 m_\chi^2 s_0 h^2}{\rho_{{\rm{cr}}}}
   \int_0^\infty\frac{dx}{x^2}\left(-\frac{dY_\chi}{dT}\Big|_{T=\frac{m_\chi}{x}}\right)\, ,
 \end{equation}
where $x=m_\chi/T$, $s_0$ the present value of the entropy density ($s_0=\frac{2\pi^2}{45}g_*T_0^3\simeq 2891.2/$cm$^3$), and $\rho_{{\rm{cr}}}$ the critical density ($\rho_{{\rm{cr}}}=1.054\times 10^{-5}h^2$GeV/cm$^3$). Eq. (\ref{DM1}) gives the observed DM abundance, that is \cite{Planck}
 \begin{equation}\label{DM2}
   \Omega_{{\rm{DM}}}h^2\Big|_{{\rm{obs}}} \ = \ 0.1188\pm 0.0010\,,
 \end{equation}
and explain the IceCube neutrino PeV signals.
The dominant contributions to DM production, induced by (\ref{4dimop}), are the {\it inverse decay} processes $\nu_\sigma+H^0\to \chi$ and $l_\sigma + H^+\to \chi$ (they are proportional to factor $|y_{\sigma \chi}|^2$  and kinematically allowed for $m_\chi> m_H+m_{\nu, l}$), and the {\it Yukawa production} processes, such as $t+{\bar t}\to {\bar \nu}_\sigma+\chi$ (they are proportional to  $|y_{\sigma \chi} y_t|^2$), where $t$ is the top quark and $y_t$ represents the Yukawa coupling constant between top quark and Higgs boson. The explicit expressions of these processes are \cite{merle}
 \begin{equation}\label{dYtot}
 \frac{dY_\chi}{dT} \ = \ \frac{dY_\chi}{dT}\Big|_{{\rm{inv.dec.}}}+\frac{dY_\chi}{dT}\Big|_{{\rm{Yuk.prod.}}}\,,
 \end{equation}
where
 \begin{eqnarray}
   \frac{dY_\chi}{dT}\Big|_{{\rm{inv.dec.}}} \ &=& \ -\frac{m_\chi^2 \Gamma_\chi}{\pi^2 s H_{{\rm{St.Cosm.}}} }\, K_1\left(\frac{m_\chi}{T}\right)\,, \label{dYinvdec} \\
   \frac{dY_\chi}{dT}\Big|_{{\rm{Yuk.prod.}}} \ &=& \ -\frac{1}{512\pi^6 s H_{{\rm{St.Cosm.}}} }\int d{\tilde s} \ \!d\Omega  \ \!\sum_\sigma
   \frac{W_{t{\bar t}\to{\bar \nu}_\sigma \chi}+2W_{t\nu_{\sigma}\to t\chi}}{\sqrt{\tilde s}}K_1\left(\frac{\sqrt{\tilde s}}{T}\right)\,. \label{dYYukprod}
 \end{eqnarray}
Here ${\tilde s}$ is the centre-of-mass energy, $W_{ij\to kl}$ are related to scattering probabilities of corresponding processes, $\Gamma_\chi$  is the interaction rate $ \Gamma_\chi = \displaystyle{ \sum_\sigma \frac{|y_{\sigma\chi}|^2}{8\pi}}\, m_\chi$, $\sigma  =  e, \mu, \tau$,
%
%
and $K_1(x)$ is the modified Bessel function of the second kind. Since $\frac{dY_\chi}{dT}\Big|_{{\rm{inv.dec.}}}$
is the dominant process (see \cite{merle} for details), one gets 
 \begin{equation}\label{DMinvdec}
   \Omega_{{\rm{DM}}}h^2|_{{\rm{inv.dec.}}} \ = \ 0.1188 \frac{\sum_\sigma |y_{\sigma \chi}|^2}{7.5\times 10^{-25}}\,.
 \end{equation}
From (\ref{DMinvdec}) it follows that the correct DM relic abundance (\ref{DM2}) occurs provided 
 \begin{equation}\label{ychi}
   \sum_{\sigma=e, \mu,\tau}|y_{\sigma\chi}|^2 \ = \ 7.5\times 10^{-25}\,.
 \end{equation}
Such a result disagrees with the value of $\sum_{\sigma=e, \mu,\tau}|y_{\alpha\chi}|^2$ needed to explain the IceCube signals. In fact, the DM lifetime $\tau_\chi=\Gamma_\chi^{-1}$ has to be larger that the age of the Universe, that is $\tau_\chi> t_U\simeq 4.35\times 10^{17}$sec. However, lower bound on DM lifetime provided by IceCube spectrum is  $\tau_\chi^b\simeq 10^{28}$sec, i.e. $\tau_\chi \gtrsim \tau_\chi^b$ \cite{merle}. Inserting (\ref{ychi}) into $\Gamma_\chi$ one gets
$\Gamma_\chi \ \simeq \ 4.5 \times 10^4 \displaystyle{\frac{m_\chi}{\text{1PeV}}}\text{sec}^{-1}$, to which corresponds $\tau_\chi \ \simeq  \ 2.2\times 10^{-5} \displaystyle{\frac{\text{1PeV}}{m_\chi}}\text{sec} \ \ll \ t_U$.
%
%
The IceCube observations require the dark matter decay lifetime of the order $\tau_\chi \simeq 10^{28}$ sec,  which implies
$\sum_{\sigma=e, \mu,\tau}| y_{\sigma\chi}|^2\ \simeq \ 10^{-58}$.
%
%
Such a value is $\sim 33$ order of magnitudes smaller than the value of $\sum_{\sigma=e,\mu, \tau}|y_{\sigma\chi}|^2\sim 10^{-25}$ that is required for explaining the DM relic abundance, cf. Eq.~(\ref{ychi}). Therefore, the IceCube events and the DM relic abundance turn out to be incompatible with the DM production if the latter is described by the renormalizable Yukawa-type operator $\overline{L}_{\alpha}H\chi$ and one works in the context of the standard cosmological model. As discussed in the main body of the paper, this discrepancy can be alleviated when Tsallis cosmology is employed without changing Yukawa-type interaction for DM.

\end{document}